\documentclass[aps,prl,preprint,superscriptaddress,nofootinbib,showkeys,floatfix]{revtex4}
\usepackage{amsmath,amssymb,graphicx}
\begin{document}
\preprint{\parbox[b]{1in}{ \hbox{\tt PNUTP-10/A11} }}

\title{Anomalous currents in dense matter under a magnetic field}


\author{Deog Ki Hong}
\email[]{dkhong@pusan.ac.kr}
\affiliation{Department of
Physics,   Pusan National University,
             Busan 609-735, Korea}
 \affiliation{Asia Pacific Center for Theoretical Physics,  POSTECH, Pohang 709-784, Korea}            


\date{\today}

\begin{abstract}
We consider fermionic dense matter under a magnetic field, where fermions couple minimally to gauge fields, and calculate anomalous currents at one loop. We find anomalous currents are spontaneously generated along the magnetic field but fermions only in the lowest Landau level contribute to anomalous currents.  We then show that there are no more corrections to the anomalous currents from two or higher loops. 
\end{abstract}


\maketitle

\section{Introduction}
\label{intro}

Fermion numbers are often not conserved, when the background fields coupled to fermions are topologically nontrivial.  One of such tantalizing effects is the non-conservation of axial fermion number under instanton backgrounds~\cite{'tHooft:1976up} and another well-known one is  the fermion number breaking in the presence of a magnetic monopole~\cite{Rubakov:1981rg,Callan:1982ac}. 
Under external background fields the energy spectrum of fermions changes and undergoes  level-crossing, which results in spontaneous creation (or annihilation) of fermions out of (or into) Dirac sea.
Direct observation of the axial fermion number non-conservation has not been made yet nonetheless, since the process, being nonperturbative, is exponentially suppressed.  However, in a recent heavy ion experiment at RHIC it has been claimed to have observed such effects~\cite{:2009uh,:2009txa}, attracting a lot of attentions. If the collision is peripheral,  a huge magnetic field is generated in the collision center, making the process easy to occur~\cite{Fukushima:2008xe}. 
  
It is well known that under a constant magnetic field, $\vec B=B\hat z$, the spectrum of charged fermions is quantized by the Landau levels, $n=0,1,2,\cdots$. At energy lower than the Landau gap, $|qB|$, where $q$ is the electric charge of fermions, the fermions move effectively one-dimensional  along the direction of magnetic field. Each fermion states are therefore characterized by two quantum numbers, $(p_{z},n)$; the momentum, $\vec p=p_{z}\hat z$,  along the field direction and the Landau level, $n$. Since three dimensional dynamics is reduced to one-dimensional by the magnetic field, the spectrum has  huge degeneracy, proportional to the Landau gap. Being one-dimensional, fermions occupy at finite density up to the Fermi points rather than Fermi surface as in two or higher dimensions: the Fermi points are given by 
$p_{F}=\pm\sqrt{\mu^{2}-2|qB|n}$, where $\mu$ is the chemical potential associated with the fermion density. Because of the Landau level degeneracy, whatever anomalous currents occur in dense matter, they are amplified by the degeneracy factor, proportional to the magnetic field, offering a window  to observe directly the topological effect of gauge fields.  
 
In this paper we carefully analyze the spontaneous generation of fermion currents in dense matter under external magnetic fields, which has been intensively studied recently, related to the relativistic heavy ion collision~\cite{Fukushima:2008xe} or magnetars~\cite{Hong:1998ka}. 

\section{Anomalous currents under external magnetic field}

When there are no external background fields coupled to fermions, the Lorentz invariance requires that the vacuum expectation value of fermion currents, $J^{\mu}=\bar\psi\gamma^{\mu}\psi$, to vanish. However, if fermions are coupled to a topologically nontrivial external field, the currents can be anomalous and the ground state may have non-vanishing vacuum expectation value of fermion currents. For instance, when charged planar fermions, confined on a plane,  are under a  (possibly position-dependent) magnetic field, $B$, perpendicular to the plane, the ground state is known to have anomalous charge density proportional to the magnetic field, since the constant magnetic flux $\Phi$ induces fermion zero modes whose number is given by the (non-negative) integer $N$ for  $\frac{e}{4\pi}\left|\Phi\right|=N+\epsilon$ with $0<\epsilon \le1$~\cite{Aharonov:1978gb,Jackiw:1984ji}; 
\begin{equation}
\left<J^0\right>_B=\pm\frac{e^2}{4\pi}B\,.\label{charge}
\end{equation}
We note that the magnetic field is a spacetime scalar for planar fermions.  Therefore, the anomalous charge density for the ground state is allowed by the spacetime symmetry. If we write it covariantly, the anomalous charge density~(\ref{charge}) becomes 
\begin{equation}
\left<J^{\mu}\right>_B=\pm\frac{e^2}{8\pi}\epsilon^{\mu\nu\rho}F_{\nu\rho}\,,
\end{equation}
which is nothing but the well-known anomalous current due to the parity anomaly or the Chern-Simons term in (2+1) dimensions~\cite{Redlich:1983kn,Niemi:1983rq,Hong:2010sb}. However, in (3+1) dimensions, the external magnetic field is a part of two-form tensor fields and thus does not generate anomalous currents because of the Lorentz symmetry. But, the situation changes for dense matter, which naturally breaks the Lorentz symmetry. In deed, it has been shown that dense matter generates anomalous currents  under a magnetic field~\cite{Fukushima:2008xe}. Here we present another derivation, but more general, of anomalous currents in dense matter under a constant magnetic field, which can be applied to dense matter with finite fermion number density and/or finite axial fermion number density. The chiral magnetic effect has been also derived in the holographic QCD~\cite{Yee:2009vw}.


The fermion propagator
 in dense matter with a chemical potential $\mu$ under a constant external magnetic field $\vec B=B\hat z$ is given by  
the Schwinger formula  as, $q$ being the electric charge of fermions,
\begin{equation}
S(x,y)={\tilde
S(x-y)}\exp\left[{iq\over2}(x-y)^{\mu}A^{\rm ext}_{\mu}(x+y)\right],
\end{equation}
with the Fourier transform of $\tilde{S}$,
\begin{equation}
{\tilde S(k)}=ie^{-k_{\perp}^2/|qB|}\sum_{n=0}^{\infty}(-1)^n
{D_n(qB, {\tilde k})\over\left[(1+i\epsilon)k_0+\mu\right]^2-k_z^2-2|qB|n},\label{spin}
\end{equation}
where $k_{\perp}$ is the 3-momentum perpendicular to the direction of
the external magnetic field, ${\tilde k}=k+(\mu,\vec 0\,)$ and 
\begin{equation}
D_n(qB,k)=2\mathord{\not\mathrel{{\tilde k}_{\shortparallel}}}\left[
P_-L_n\left({2k_{\perp}^2\over
|qB|}\right)-P_+L_{n-1}\left({2k_{\perp}^2\over |qB|}\right) \right]
+4 \mathord{\not\mathrel{k_{\perp}}}
 L_{n-1}^1\left({2k_{\perp}^2\over |qB|}\right)\,,\label{propagator}
\end{equation}
where ${\tilde k}_{\shortparallel}=(k_0+\mu,k_z)=k_{\shortparallel}+(\mu,\vec 0\,)$. 
$L_n^{\alpha}$ are the associated Laguerre polynomials and $P_{+}$
($P_-$)
is the projection operator which projects out the fermions of
spin (anti-) parallel to the magnetic field direction.
For $\vec B=B\hat z$,
$2P_{\pm}=1\pm i\gamma^1\gamma^2 {\rm sign} (qB)$.

For a later purpose we decompose the fermion propagator into chiral basis:
\begin{equation}
{\tilde S(k)}=\frac{1+\gamma^5}{2}{\tilde S(k)}+\frac{1-\gamma^5}{2}{\tilde S(k)}
={\tilde S(k)}_L+{\tilde S(k)}_R\,.
\end{equation}
Now we assume that the chemical potentials for the left-handed fermions  and the right-handed fermions are different, $\mu_L\ne\mu_R$. The fermion number chemical potential is then $\mu=(\mu_L+\mu_R)/2$ and the axial chemical potential $\mu_A=(\mu_L-\mu_R)/2$. 
The anomalous current for the left-handed fermions is given at one-loop by 
\begin{eqnarray}
\Delta_L^{\alpha}(\mu_L)\equiv\left<\bar\psi_L\gamma^{\alpha}\psi_L\right>=-\int\frac{{\rm d}^4k}{(2\pi)^4}{\rm Tr}\left(\gamma^{\alpha}{\tilde S(k)}_L\right)\,.\label{anomalous}
\end{eqnarray}
The integration~(\ref{anomalous}) is in general ultra-violet (UV) divergent and needs to be regularized. However the UV divergence is due to the vacuum contribution and is independent of the chemical potential. In other words, the matter contribution is finite and can be written as 
\begin{equation}
\Delta^{\alpha}_{\rm mat}(\mu_L,B)\equiv \Delta^{\alpha}(\mu_L, B)-\Delta^{\alpha}(0,B)=\int_0^{\mu_L}\!{\rm d}\mu^{\prime}\,\frac{\partial}{\partial\mu^{\prime}}\Delta^{\alpha}(\mu^{\prime},B)\,,\label{matter}
\end{equation}
where we have subtracted out the vacuum contribution $\Delta^{\alpha}(0,B)$.

The integration (\ref{matter}) is finite and explicitly calculable. 
To this end, we change the integration variables, which is well defined for a finite integration:
\begin{equation}
k^{\alpha}\longrightarrow {k^{\prime}}^{\alpha}=k^{\alpha}+u^{\alpha}\mu\,,\quad u^{\alpha}=(1,{\vec 0}\,)\,
\end{equation}
and rewrite the fermion propagator as 
\begin{equation}
{\tilde S}_L(k-u\mu_L\,;\mu_L)=ie^{-k_{\perp}^2/|qB|}\sum_{n=0}^{\infty}(-1)^n\frac{D_n(qB,k)}{k^2_{\shortparallel}-2|qB|n+i\epsilon\, k_0(k_0-\mu_L)}\,.
\end{equation}
Differentiating with respect to the chemical potential, we get 
\begin{equation}
\frac{\partial}{\partial \mu}{\tilde S}(k-u\mu\,;\mu)=ie^{-k_{\perp}^2/|qB|}\sum_{n=0}^{\infty}(-1)^n D_n(qB,k)\,
2\pi i\,\delta\!\left(k_{\shortparallel}^2-2|qB|n\right)\cdot\delta (k^0-\mu)\,,
\label{diff}
\end{equation}
which clearly shows that the momentum integration in~(\ref{matter}) has supports only from the Fermi points and on-shell, rendering the total integration to be finite. As we can see from the expression~(\ref{spin}) of the propagator of fermions, the fermions in each Landau level except the lowest Landau level (LLL) have both spins, parallel and anti-parallel to the external magnetic field. 

\begin{figure}[tbh]
	\centering
	\includegraphics[width=0.8\textwidth]{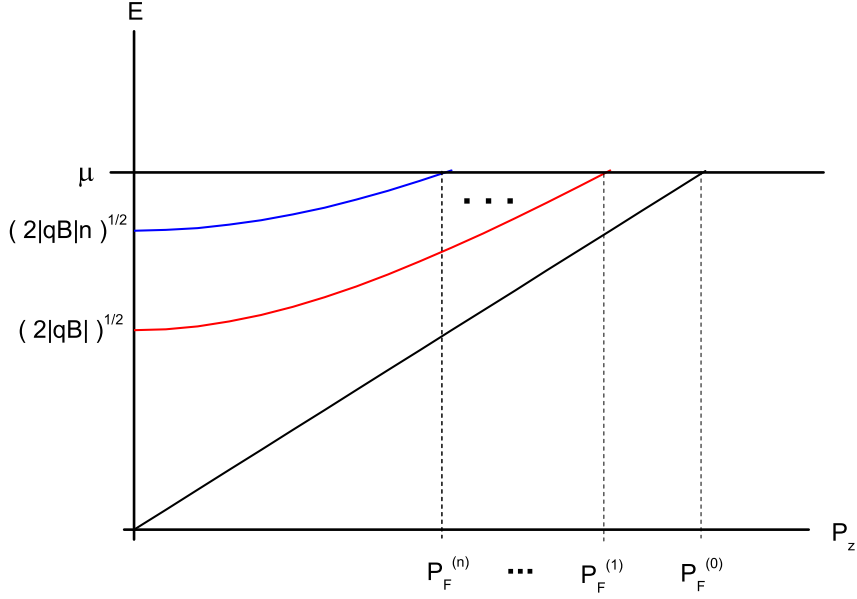}%
		\caption{\label{fermi_m}Fermi momenta in different Landau levels}
\end{figure}

Using a normalization condition for the Laguerre polynomials,  
\begin{equation}
\int_0^{\infty}{\rm d}x\,e^{-x}L_n(2x)=(-1)^n\,,
\end{equation}
we  integrate over the perpendicular momentum, $\vec k_{\perp}$, to get, after taking the trace over gamma matrices, 
\begin{equation}
\Delta_{\rm mat}^{\alpha}(\mu_L,B)=|qB|\left[\Gamma_L^{\alpha\beta}I_{\beta}^{(0)}(\mu_L,B)+
2g_{\shortparallel}^{\alpha\beta}\sum_{n=1}I^{(n)}_{\beta}(\mu_L,B)\right]\,,\label{sum}
\end{equation}
where $g_{\shortparallel}^{\alpha\beta}={\rm diag}\,(1,0,0,-1)$, $\Gamma_L^{\alpha\beta}=\epsilon^{\alpha\beta12}\,{\rm sign}(qB)+g_{\shortparallel}^{\alpha\beta}$ and for $n=0,1,2,\cdots$
\begin{equation}
I^{(n)\beta}(\mu,B)=\int_0^{\mu}{\rm d}{\mu^{\prime}}\int\frac{{\rm d}^2k_{\shortparallel}}{(2\pi)^2}k_{\shortparallel}^{\beta}\delta\!\left(k_{\shortparallel}^2-2|qB|n\right)\cdot\delta (k^0-\mu^{\prime})=\frac{1}{4\pi^2}\,p_F^{(n)}(\mu,B)\,\delta^{\beta0}\,.\label{int}
\end{equation}
The Fermi momentum at the $n$-th Landau level is given as (see Fig.~\ref{fermi_m})
\begin{equation}
p_F^{(n)}(\mu,B)= \begin{cases}
\sqrt{\mu^2-2|qB|n},    &\text{if $\mu>2|qB|n $;}\\
0,                                                                &\text{otherwise.}
\end{cases}
\end{equation}
We see that only the LLL fermions contribute to the third component of anomalous current
while fermions in the  levels up to $n<\mu/2|qB| $ do contribute to the charge density:
\begin{eqnarray}
\Delta_{\rm mat}^3(\mu_L,B)&=&\frac{qB}{4\pi^2}\,\mu_L\,,\label{current}\\
\Delta_{\rm mat}^0(\mu_L,B)&=&\frac{|qB|}{4\pi^2}\left[\mu_L+2\sum_{n=1}p_F^{(n)}(\mu_L,B)\right]\,,\label{density}
\end{eqnarray}
where the factor 2 in the parenthesis of~(\ref{density}) is the spin degeneracy of higher Landau levels.   
Since each Landau level has a degeneracy factor $|qB|/4\pi$, we see that the charge density in Eq.~(\ref{density}) is precisely the total number density of left-handed fermions, $\Delta^0_{\rm mat}(\mu_L,B)=n_L$. 

Now, we note that the vacuum contribution to the anomalous current~(\ref{anomalous}) is zero for the LLL fermions because of the residual Lorentz symmetry $SO(1,1)$ along time and the B-field direction;
\begin{equation}
\Delta_L^{\alpha}(0,B)=2\Gamma_L^{\alpha\beta}\int\frac{\!{\rm d}^4k}{(2\pi)^4}\,\frac{{ k}_{\shortparallel\beta}}{k_{\shortparallel}^2+i\epsilon}\, e^{-k_{\perp}^2/|qB|}=0\,.
\end{equation} 
Therefore we find $\Delta_L^{\alpha}(\mu_L,B)=\Delta^{\alpha}_{\rm mat}(\mu_L,B)$, which holds for the chemical potential larger than the Landau gap $|qB|$ as well.  The anomalous current generated in (chiral) dense matter therefore becomes 
\begin{equation}
\Delta_L^{\alpha}(\mu_L,B)=\delta^{\alpha3}\,{\rm sign}(qB)\frac{|qB|}{4\pi^2}\mu_L+\delta^{\alpha0}\,n_L\,.
\label{anomalous3}
\end{equation}
Similarly for the right-handed fermions, the anomalous current is given   as 
\begin{equation}
\Delta_R^{\alpha}(\mu_R,B)=-\delta^{\alpha3}\,{\rm sign}(qB)\frac{|qB|}{4\pi^2}\mu_R+\delta^{\alpha0}\,n_R\,.
\label{anomalous2}
\end{equation}
Therefore in terms of electric (axial) vector currents we find, agreeing with the previous calculation~\cite{Fukushima:2008xe}, 
\begin{eqnarray}
J_V^{\alpha}\equiv q\left(\Delta_L^{\alpha}+\Delta_R^{\alpha}\right)&=&\delta^{\alpha3}\,\frac{q^2B}{2\pi^2}\mu_A+\delta^{\alpha0}\,q\,n\,,\label{vector}\\
J_A^{\alpha}\equiv q\left(\Delta_L^{\alpha}-\Delta_R^{\alpha}\right)&=&\delta^{\alpha3}\,\frac{q^2B}{2\pi^2}\mu+\delta^{\alpha0}\,q\,n_A\,,\label{axial}
\end{eqnarray}
where the fermion number density $n=n_L+n_R$ and the axial fermion number density $n_A=n_L-n_R$~\footnote{The anomalous currents could have additional contributions if the fermions are not minimally coupled or their propagator takes a non-standard form~\cite{Gorbar:2009bm,Fukushima:2010zza}.}.

\section{No more corrections to anomalous currents}
In the previous section we have calculated the one-loop  anomalous currents in dense matter. Now, we show that the results are in fact exact in all orders. Namely the results are not subject to any higher order corrections. In both results~(\ref{vector}) and  (\ref{axial}) the second terms are nothing but  the volume inside the Fermi points, namely the fermion number density, which is not renormalized by interactions and does not get any corrections by the Luttinger theorem~\cite{Luttinger:1960zz}. So, we need only to show that the first terms in~(\ref{vector}) and  (\ref{axial}) are not subject to higher order corrections. We first consider the anomalous vector currents, setting $\mu_A\ne0$ but $\mu=0$ without loss of generality. The proof for the anomalous axial current follows immediately, once the proof for the no-more corrections to the vector current is established, because  
the diagrams are same except  the one un-contracted photon line is now replaced by an auxiliary axial $U(1)$ gauge field, which couples to the axial $U(1)$ current.

\begin{figure}[tbh]
	\centering
	\includegraphics[width=0.8\textwidth]{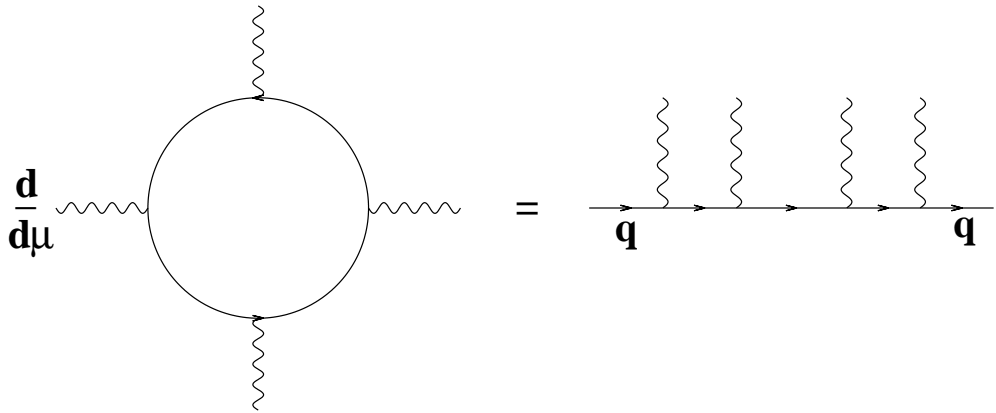}%
		\caption{\label{vertex}One of the vertex correction diagrams before contracting gauge fields.}
\end{figure}

We compute the full contributions to the anomalous current by differentiating the effective action with respect to the photon field; 
\begin{equation}
\left<J^{\alpha}\right>=\left.\frac{\delta \Gamma(A,G \,; \mu)}{\delta A_{\alpha}}\right|_{A=0=G}\,,
\end{equation}
where $\Gamma$ is the effective action for photons ($A$) and gluons ($G$). The effective action is obtained by two steps~\cite{Coleman:1985zi}. First we integrate out the fermions to get the one-loop $n$-point vertex functions $\Gamma^{(n)}$ for photons and gluons. We then contract all the photon (or gluon) lines with other photon (or gluon) lines, which do not have to end on the same fermion loop,  except one photon line to get a full photon-tadpole diagram.\footnote{Furry's theorem forbids odd number of photon external lines in a theory with the charge conjugation symmetry.  However, in dense matter the charge conjugation symmetry is explicitly broken by the chemical potential $\mu$, which is odd under the charge conjugation. It is therefore obvious that the anomalous current, if non-vanishing, should be proportional to the chemical potential.}  As explained in the previous section the matter contribution to the anomalous current is finite and can be written as 
\begin{equation}
\left<J^{\alpha}\right>_{{\rm mat}}=\frac{\delta\Gamma_{\rm mat}({\cal A})}{\delta A_{\alpha}}=\frac{\delta}{\delta A_{\alpha}}\int_{0}^{\mu}\!{\rm d}{\mu^{\prime}}\,\frac{\partial}{\partial \mu^{\prime}}\Gamma({\cal A} \,; \mu^{\prime})\,,
\end{equation}
where ${\cal A}=A$ or $G$ are taken to be zero after the differentiation.  
For the one-loop $n$-point vertex function there are $n$ fermion propagators in the fermion loop. If we take a derivative with respect to the chemical potential, after shifting the momentum as in the previous section,  one of the propagator gives a delta function,  enforcing the on-shell condition:
\begin{equation}
\frac{\partial}{\partial \mu}{\rm Tr}\!\left[\not\!\!{\cal A}S(k_{1})\!\not\!\!{\cal A}\cdots \not\!\!{\cal A}S(k_{n})
\right]=\!\sum_{l=1}^{n}{\rm Tr}\!\left[\not\!\!{\cal A}S(k_{1})\cdots\not\!\!{\cal A}\!\not\!\!k_{l\shortparallel}\! \not\!\!{\cal A}\cdots\not\!\!{\cal A}S(k_{n})\right]2\pi i\,\delta (k_{l\shortparallel}^2)\delta (k_{l}^0-\mu)\,,
\end{equation}
where we have suppressed the other terms present in the fermion propagator.
Therefore the derivative of the $n$-point one-loop vertex function with respect to the chemical potential gives the tree-level on-shell fermion propagator with $n$ gauge fields attached (See Fig.~\ref{vertex}), whose total momentum gives zero. 
Now, if we contract all the photon or gluon lines, which not necessarily end on the  same fermion loop or on the tree-level on-shell fermion propagator,  except one photon line, $\Gamma_{\rm mat}({\cal A})$ gives the full one-point vertex function at zero momentum with or without wave-function renormalizations to the external fermion lines. The un-contracted photon line can end on either the tree propagator or a fermion loop~\footnote{When $\mu=0$, this way of generating vertex correction will miss diagrams where the external photon line ends on the internal fermion loop because of Furry's theorem~\cite{Itzykson:1980rh}. But, in dense matter it does not apply.}.  
Since the wave-function renormalization vanishes for the external lines, which are on-shell, the vertex function obtained by taking the derivative of the effective action with respect to the chemical potential gives the full vertex function at zero momentum transfer. 
Because of the non-renormalization of electric charge or by the Ademollo-Gatto theorem~\cite{Ademollo:1964sr} there should be no vertex correction at zero momentum from one-loop or higher.   Furthermore, since the vacuum ($\mu=0$) contribution to the anomalous current vanishes because of the residual Lorentz symmetry,  the one-loop result is therefore exact and the anomalous current does not get any higher loop corrections. 

In conclusion, we have found the ground state of dense matter under a magnetic field has anomalous currents along the magnetic field direction. Only fermions in the lowest Landau level contribute to the anomalous currents along the magnetic field, though the fermions in higher levels do contribute to the total number density. The fermions in the higher Landau levels do not contribute to the anomalous currents because of spin degeneracy. Furthermore, we show that there are no corrections to the anomalous currents from two or higher loops.

\subsection{Acknowledgements}
The author  thanks C. Ahn, P. Fulde, C. Lee, and I. Shovkovy for useful discussions and helpful comments. This work was supported by the Korea Research
Foundation Grant funded by the Korean Government (KRF-2008-341-C00008).

\end{document}